\documentclass[aps,prl,reprint,showpacs,superscriptaddress,nobibnotes,nofootinbib]{revtex4-2}
\usepackage{graphicx}
\usepackage{amssymb} 
\usepackage{amsmath}
\usepackage{pifont}
\usepackage{enumerate}
\usepackage [colorlinks=true, linkcolor=blue, citecolor= blue, urlcolor=blue,]{hyperref}
\begin{document}

\title{Precision magnetometry at cryogenic temperatures with gaseous $^3\text{He}$ NMR probes}
\author{P. Bl\"umler}
\affiliation{Institut f\"{u}r Physik, Johannes Gutenberg-Universit\"{a}t, 
	55128 Mainz, Germany}
\author{M. Fertl}
\affiliation{Institut f\"{u}r Physik, Johannes Gutenberg-Universit\"{a}t, 
	55128 Mainz, Germany}
\author{H.-J. Grafe}
\affiliation{Leibniz Institute for Solid State and Materials Research (IFW), 01069 Dresden, Germany}
\author{R. Graf}
\affiliation{MPI for Polymer Research, 55128 Mainz, Germany}
\author{W. Heil}
\email{Corresponding author: wheil@uni-mainz.de}
\affiliation{Institut f\"{u}r Physik, Johannes Gutenberg-Universit\"{a}t, 
	55128 Mainz, Germany}

\date{\today}
\begin{abstract}
We report on compact, gaseous $^3\text{He}$ NMR probes for precision magnetometry of magnetic fields $B > 0.1$ T in the temperature range from ambient temperatures down to 4 Kelvin. The gas is polarized at thermal equilibrium under pressures up to 100 bar to provide a high nuclear spin density. In order to achieve sensor readout rates of $\mathcal{O}$ (Hz), paramagnetic substances and/or silica gel with high specific surface area were added to reduce the otherwise long $T_1$ relaxation time of pure $^3\text{He}$ gas to reach thermal polarization equilibrium. Sensitivity limits, which cover the range from $10^{-11} < \delta B/B < 10^{-7}$, are accessible in a single-pulse NMR measurement and can be further improved through signal averaging in accumulated NMR scans.
\end{abstract}


\maketitle


{\Large\bf 1. Introduction}\\

Ultra-sensitive measurements and monitoring of high magnetic fields $(B > 0.1~ \text{T})$ are of great interest for different fields of physics and applied research, ranging from accelerator science (e.g. BNL/FNAL, Muon $g-2$ experiment \cite{Fei97, Bennett06, Swanson24}), to mass spectroscopy \cite{Blaum06} and practical applications such as shimming procedures for permanent and superconducting magnets \cite{Fuks92} or monitoring field dynamics in MRI systems \cite{Zanche08}.
At high magnetic fields NMR probes (Nuclear Magnetic Resonance) give the highest accuracy \cite{Bottura02}. Due to their capability of determining absolute field values, they are commonly used for the calibration of magnetic sensors which are based on other physical principles, e.g., Hall effect sensors. Until recently and almost exclusively, small liquid samples (e.g. water or cyclohexane) served as NMR-samples for this purpose. For example, X. Fei et al. \cite{Fei97}
tested precision NMR probes to characterize the magnetic field of a conventional magnetic resonance imaging (MRI) magnet $(1.5~\text{T})$, reaching an  accuracy of $\approx 4\times 10^{-8}$ in the determination of the absolute magnetic field. In this work, the essential gain in sensitivity was reached using a susceptibility-matched spherical water sample. Already in X. Fei et al. \cite{Fei97} and somewhat earlier by J.L. Flowers et al. \cite{Flowers93}, it is stated that optically pumped polarized $^3\text{He}$ $(I=1/2)$ may be a more appropriate NMR sample to increase the sensitivity limits. In Ref. \cite{Nikiel14}, we describe a $^3\text{He}$ magnetometer capable of measuring magnetic fields of the order of Tesla to a relative precision of better than $10^{-12}$. The $^3\text{He}$ gas is spin polarized {\it in situ} using a new, nonstandard variant of metastability exchange optical pumping (MEOP). In this context, $^3\text{He}$ nuclear magnetic resonance probes have been proposed as a new standard for absolute magnetometry \cite{Nikiel14, Farooq20}. This requires a high accuracy value for the $^3\text{He}$ nuclear magnetic moment, which, however, has so far been determined only indirectly, based on water NMR probes \cite{Flowers93}. Recently, a direct measurement of the $^3\text{He}$ nuclear magnetic moment was performed by investigating the hyperfine structure of a single $^3\text{He}^{+}$ ion in a Penning trap \cite{Schneider22}, thus improving the precision of the shielded magnetic moment by an order of magnitude, i.e., $\mu_\text{He}/h = -16.217050033(14) \text{MHz/T}$.\\
\indent The use of gaseous $^3\text{He}$ NMR magnetometers can also be extended to cryogenic temperatures without loss of accuracy. Here, most NMR sensors are not suitable or at least have limitations because specialized spin samples such as metallic $^{27}\text{Al-powder}$ \cite{Borovikov01} or $^1\text{H}$-NMR on samples composed of acetone~/~ethanol mixtures and water, of course, are all in the solid state and the dipolar coupling of the spins, which is usually averaged out in NMR spectra of solutions or gaseous samples, leads to substantial broadening of the spectra along with severely degraded precision \cite{Vleck48}. 
As superconducting magnets become more common, there is an enormous demand for compact, reliable, and easy-to-handle cryogenic NMR probes. $^3\text{He}$ magnetometers based on free spin precession after resonant radio frequency pulse excitation would, for the first time, provide a field sensor capable of operating across the full temperature range of interest, from 4 Kelvin to room temperature. 
To perform precision magnetometry with $^3\text{He}$ NMR samples at low temperatures, key limitations must be addressed and overcome:\\
\indent When going to cryogenic temperatures, MEOP becomes ineffective and slow since the orientation transfer via metastability exchange (ME) collisions becomes less efficient. For a given $2^3\text{S}$ density, the ME collision rate is reduced by a factor of about 30 at $\approx \text{4~K}$ \cite{Colegrove64, Fritz68}. Taking into account that a large number of ME collisions, proportional to $B^2$, are required to transfer a single unit of angular momentum between the ground-state and the $2^3\text{S}$ state atoms \cite{Abboud04}, the polarization build-up may take several minutes. Although somewhat challenging to integrate a device for optical pumping in a low temperature setting, MEOP at cryogenic temperatures has been successfully demonstrated in references \cite{Barbe75, Himbert83}.  Most experiments were performed with Pyrex spherical cells $(\oslash \approx 3~\text{cm})$ at a $^3\text{He}$ gas filling pressure of a few millibars. Another approach, as described in~\cite{Leduc84}, utilizes a polarization transfer technique in which hyperpolarized $^3\text{He}$ gas from a room temperature MEOP reservoir diffuses through a connecting tube to a second cell immersed in a liquid helium bath. In this way, ground-state nuclear polarizations of up to $50\%$ and number densities of order $10^{18}$~$\text{cm}^{-3}$ have been produced. The main drawback of this method is that the polarization at low temperature increases slowly and reaches a maximum after a delay of the order of 1 hour when the upper cell is continuously pumped.\\
\indent Another coupled reservoir-bulb system using thermally polarized $^3\text{He}$ was realized by X. Fan et al. \cite{Fan19}. It consists of a large room temperature reservoir (1.2~liters at $\approx 1\text{bar}$ 
 $^3\text{He}$) connected to the NMR bulb (0.5~$\text{cm}^3$) at 4.2~K via a capillary tube to ensure high density of the $^3\text{He}$ gas. Although the longitudinal relaxation time constant $T_1$ to reach thermal polarization equilibrium was long, i.e. 364(31)~s, NMR measurements with a recovery time of 20~s were sufficient to give a signal-to-noise ratio of about 10.\\
 \indent In the following, we report on the development of a $^3\text{He}$ magnetometer capable of covering the entire low-temperature range $4~\text{K} < T < 300~\text{K}$ for magnetic fields $B > 0.1~\text{T}$. As in~\cite{Fan19}, we use thermally polarized $^3\text{He}$ NMR probes, but instead of a coupled reservoir-bulb system, sealed high pressure samples of $^3\text{He}$ are utilized to provide a high spin density. The main drawback in using thermally polarized $^3\text{He}$ is the long $T_1$-time of minutes to hours to reach thermal polarization equilibrium, which gives rise to slow sensor read-out rates and was hitherto the main argument against use for magnetometry. We successfully introduced three strategies to reduce $T_1$ over wide temperature ranges, so this obstacle has also been mastered. \\

{\Large\bf 2. Methodology background}\\

The basic principle of NMR is to polarize the magnetic moments $\mu_I$ of nuclei (here: $^3\text{He}$) along the axis of the respective magnetic field ($z$-axis), and then to tip them synchronously away from that axis towards the transverse $x-y$ plane by applying a short resonant radio frequency (rf) pulse. Subsequently, the free, coherent precession of the nuclear magnetic moments around the field direction with the Larmor frequency
\begin{equation} \label{equ:1} 
\omega_0/2\pi=\gamma_\text{He}\cdot \lvert{\bf B}\rvert,
\end{equation}

 \noindent is detected by means of a receiver coil of quality factor $Q$, which is tuned to resonate at the Larmor frequency. The proportionality constant $\gamma_\text{He}=2\cdot (\mu_\text{He}/h)$ is called the gyromagnetic ratio and is the factor used to transform the measured NMR resonant frequency into a value for the magnetic flux density $\lvert{\bf B}\rvert$. The loss of coherence of the NMR-signal called Free Induction Decay (FID) is usually characterized by an exponential decay with a time constant $T_2^*$, as field inhomogeneities over the sample’s volume cause the precessing nuclear spins to get out of phase with each other. 
When the $^3\text{He}$ gas pressure and the magnetic field are sufficiently high, the spin precession time becomes short relative to the diffusion time across the inhomogeneous field, $\Delta B$, inside the sample’s volume, preventing motional narrowing \cite{Cates88}. In this case, the transverse decay constant of the spin precession signal can be deduced to be \cite{Slichter90}
\begin{equation} \label{equ:2} 
\frac{1}{T_2^*}=\frac{1}{T_2}+2\pi\gamma_\text{He}\Delta B\approx 2\pi\gamma_\text{He}\Delta B ~.   
\end{equation}

This time constant can range from milliseconds to fraction of a second, depending on the nature of the sample and the homogeneity of the magnetic field.  Here, we have neglected the spin-spin relaxation time constant $T_2$ which contributes even when the external magnetic field is perfectly homogeneous $(\Delta B = 0)$ and which was determined to be $T_2\approx 2.6~\text{s}$ from a Carr-Purcell-Meiboom-Gill (CPMG) NMR pulse sequence \cite{Ziegler99, Fan19}. The usual formula for the signal-to-noise ratio ($\mathit{SNR}$) available after a $\pi$/2 pulse is given by \cite{Abragam61,Hoult76}
\begin{equation} \label{equ:3} 
\mathit{SNR}\,=\,K\cdot\eta \left (\,\frac{\mu_0\, Q\,\omega_0\, V_\text{c}}{4F\,k_\text{B}\,T\,f_\text{BW}}\right )^{1/2}\cdot M_0,
\end{equation}

\noindent where $K$ is a numerical factor $(\simeq 1)$ dependent on the geometry of the receiving coil, $\eta$ is the “filling factor,” that is, a measure of
the fraction of the coil volume $(V_\text{c})$ occupied by the sample $(V_\text{s})$, $\mu_0$ is the permeability of the free space, $F$ is the noise figure of the spectrometer, $k_\text{B}T$ is the thermal energy of the sample/coil at temperature $T$ with the Boltzmann constant $k_\text{B}$, and $f_\text{BW} = 1/\Delta t$ is the measurement bandwidth with $\Delta t$ being the sampling interval. The magnetization $M_0$ of a thermally polarized $^3\text{He}$ sample is given by
\begin{equation} \label{equ:4} 
M_0=N\cdot \mu_\text{He}\cdot \tanh\left (\frac{\mu_\text{He}\cdot B}{k_\text{B}T}\right ) \,\approx\, N\mu_\text{He}^2 \frac{B}{k_\text{B}T},
\end{equation}

\noindent where $N$ is the number density of spins at resonance. The hyperbolic tangent factor is the net fraction of the spins that are
thermally aligned. The approximation to the right in Eq. (\ref{equ:4}) suffices in all the cases considered here. 
From Eqs.(\ref{equ:3},\ref{equ:4}) we deduce a  dependence of the $\mathit{SNR}$, resulting in a gain factor of $\simeq 600$ in measurement sensitivity when going from room temperature to 4.2\,K.  
However, one has to take into account the factor $F$ in Eq.(\ref{equ:3}) that comprises all additional noise sources to and from the spectrometer. Often it is convenient to express the noise performance in terms of the excess noise temperature $T_\text{e}$ \cite{Doty88}. This is the amount by which the source temperature $T$ would have to be raised to produce the same noise output from an ideal noise-free amplifier of the same gain and $F$, $T$ and $T_\text{e}$ are related by $F= 1+T_\text{e}/T$. The typical noise temperature of the receiver (preamplifier) at room temperature is $T_\text{e} = 50~\text{K}-200~\text{K}$ \cite{Hill68}, therefore, if no other measures are taken to suppress this excess noise, we expect the temperature dependence of the $\mathit{SNR}$ to be $\propto 1/T$. In an NMR magnetometry context, the Cram\'{e}r-Rao Lower Bound (CRLB) is the lower limit of the uncertainty on the NMR resonance value in the presence of Gaussian noise \cite{Kay93}. The lower limit on the variance $\sigma_f^2$  for the frequency estimation of an exponentially damped sinusoidal signal is given by
\begin{equation} \label{equ:5} 
\sigma_f^2\,\geq\, \frac{12}{(2\pi)^2\, \mathit{SNR}^2 \,f_\text{BW}\,t^3}\cdot C(\kappa),
\end{equation}

\noindent where $t$ is the acquisition time and $C(\kappa)$ describes the effect of exponential damping of the signal amplitude with $T_2^*$ \cite{Gemmel10} and is given by 
\begin{equation} \label{equ:6} 
C(\kappa =t/T_2^*)=\frac{2}{3}\kappa^3\frac{\left (1-\text{exp}(-2\kappa)\right )}{\left (1-\text{exp}(-2\kappa)\right )^2-4\kappa^2\text{exp}(-2\kappa)}.
\end{equation}

Finally, the sensitivity $\delta B$ on the respective magnetic field $B$ is given by $\delta B\leq\sigma_f/\gamma_\text{He}$.
When the acquisition time is $t\ll T_2^*$, the signal is effectively undamped. The precision follows the $t^{-3/2}$ power law and the limit $\sqrt{C}\to 1$  is obtained. For observation times much longer than $T_2^*$, $\sqrt{C}\propto (t/T_2^*)^{3/2}$ and $\delta B$ are independent of $t$. This describes a situation where no additional information about the resonance frequency is obtained from a longer observation. In practice, the acquisition time is set to $t\approx 3T_2^*$ . Often the FID envelope may differ from a pure exponential function corresponding to suboptimal magnetic field homogeneity conditions, e.g., due to imperfect shimming of the magnet. In this case, the CRLB derived for $\sigma_f$  or $\delta B$ can only be seen as an approximate value.\\ 

{\Large\bf 3.~Experimental}\\

{\bf 3.1~Provision of high-pressure gas samples} \\

\begin{figure}[h!]
    \centering
    \includegraphics[width=0.48\textwidth]{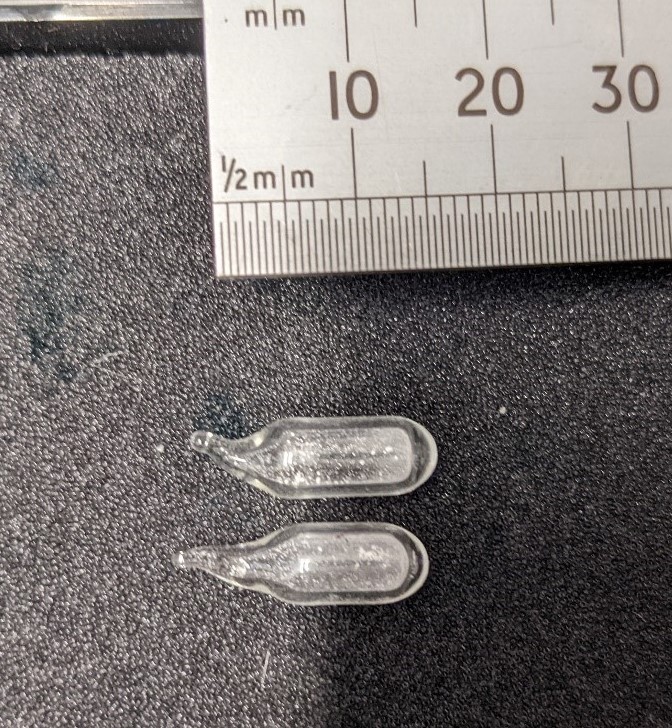}
    \caption{Photograph of two fused silica sample cells sealed off under vacuum. The white powder inside is strontium peroxide to release oxygen; see text for details. } 
    \label{fig:1}
\end{figure}

The main problem in making high-pressure gas samples sealed in glass or fused silica cells is the fact that the cells cannot be flame-sealed with higher internal than external pressures, which limits the inner pressure to ambient pressure. To avoid this problem, we used the selective permeation of helium through quartz to fill the cell with a defined helium pressure, a technique that was already utilized in an earlier publication on this topic \cite{Maul16}. In Fig.~\ref{fig:1} we show two of our sealed-off sample cells, which were made from 6 mm quartz tubes with an inner diameter of 4 mm and which were prepared with a sealing neck for a length of ca. 10 mm and hence an inner volume of $V_\text{s} = 0.126(10)~\text{cm}^3$ after the cell was sealed off.  The bursting pressure $p_\text{b}$ of quartz tubes follows the Lam\'{e} formula when cylinders (quartz tubes) are subjected to internal pressure:~$p_\text{b}=S\cdot (R^2-1)/(R^2+1)$ \cite{Kyi50}.  With $R$ being the ratio of external to internal diameter and $S$ the tensile strength of the tube material, we obtain $p_\text{b}\simeq 200~\text{bar}$ for $R=1.5$ and $S_\text{quartz}= 480~\text{bar}$. For short cylinders with closed ends, as in our case, this value is still increasing slightly, so the safety margin for filling pressures is $p\leq 100~\text{bar}$.  

\begin{figure}
    \centering
    \includegraphics[width=0.48\textwidth]{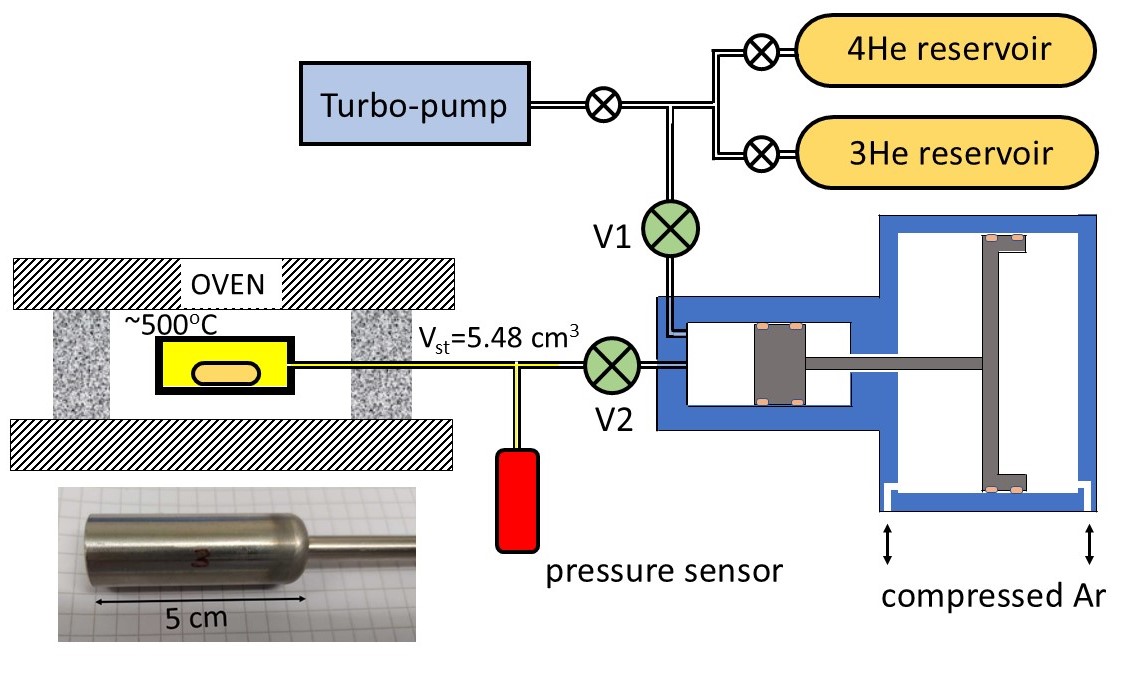}
    \caption{Sketch of the $^3\text{He}$ filling station. Tubes, fittings, adapters and valves are designed accordingly for high-pressure use (Swagelok). The inset shows a photograph of the hollow stainless-steel cylinder which is welded to a connecting tube, and in which a sealed off quartz cell was inserted beforehand. After the filling procedure, the cell is accessed by opening the cylinder at one end by means of an iron saw. See text for more details.} 
    \label{fig:2}
\end{figure}

The apparatus used to fill the quartz cells with helium via permeation is shown in Fig.~\ref{fig:2}. By means of a turbo pumping station all connecting pipes $(\oslash_\text{I}= 2~\text{mm})$ including the compressor unit and the bake-out section with the stainless steel cylinder at its very end housing the quartz cell, can be evacuated. Two gas bottles $(V = 1~\text{liter})$ filled with $^3\text{He}$ respectively $^4\text{He}$ at a maximum pressure of $3~\text{bar}$ provide the gas supply. The pneumatically driven compressor is a home-made unit and, with its stroke volume of $160~\text{cm}^3$, has an effective compression factor of about 20. The two bellows-sealed valves V1 and V2 (SS-6BK-MM from Swagelok) are helium tight with a dead volume of only $1.6~\text{cm}^3$ and can be used up to a working pressure of $100~\text{bar}$. If necessary, the compressor can be activated several times to compress sufficient amount of gas from the gas bottle into the intentionally small storage volume of $V_\text{st}=5.48(7)~\text{cm}^3$ downstream of valve V2. Alternatively, the compressor can be used as a pump to return a high percentage of the remaining gas in $V_\text{st}$ to one of the gas bottles after the filling process of the cell has been completed. This mainly concerns the rather expensive $^3\text{He}$ gas. The pressure is determined via an electronic pressure sensor (Autosen AP018) with a measuring range from 0 to $100~\text{bar}$ and a resolution of $10~\text{mbar}$. 

\begin{figure}
    \centering
    \includegraphics[width=0.48\textwidth]{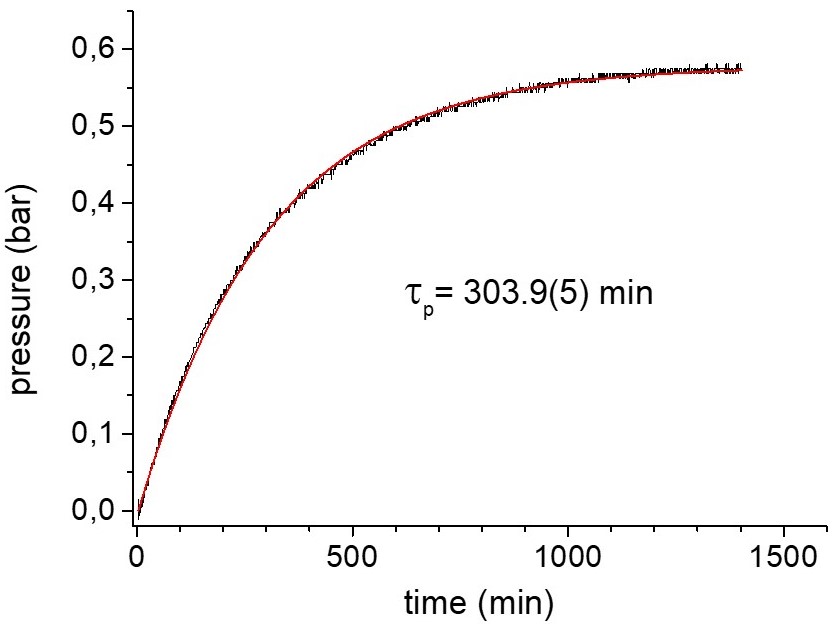}
    \caption{Build-up of the $^4\text{He}$ pressure due to gas permeability into the small storage volume of $V_\text{st}=5.48(7) \text{cm}^3$, which is kept at $\simeq500^\circ\text{C}$ and which houses a quartz cell previously filled with $^4\text{He}$. The residual gas from the filling process was pumped out beforehand. An exponential fit (red curve) gives the characteristic permeation time constant of $\tau_\text{p}=303.9(5)\text{min}$, i.e., about 5~h.  } 
    \label{fig:3}
\end{figure}

In order to determine the characteristic permeation time constant $\tau_\text{p}$ for this type of cells, $V_\text{st}$ was filled with $^4\text{He}$ and the front part of the connecting tube with the stainless steel cylinder was baked out in a hinged tube furnace at $~\simeq~500^\circ\text{C}$ (see Fig. \ref{fig:2}). After about 30 hours, the residual gas was pumped out and, with V2 closed again, the increase in pressure due to repermeation was monitored, which should show exponential saturation behavior according to $p(t)=p_0[ 1-\text{exp}(-t/\tau_\text{p})]$. Based on the measured time constant of $\tau_\text{p}\simeq~5~\text{h}$ (see Fig. \ref{fig:3}), we have set the total time required to fill each of our cells at $t_\text{f}\simeq~5\cdot\tau_\text{p}$, that is, approximately one day. The same time span during which pressure equilibrium is reached was also used for $^3\text{He}$ as a conservative limit, since the lighter isotope should permeate faster \cite{Jones53}. According to Swets et al. \cite{Swets61}, the permeability of helium through quartz at $500^o\text{C}$ is $P_{500} = 2.74(23)\times10^{12}~\text{atoms/cm/s}$ at a pressure difference $\Delta p = 1~\text{bar}~(10^5~\text{Pa})$, so the expected permeation time constant can be deduced as 
\begin{equation} \label{equ:7} 
\tau_\text{p}\,=\,\frac{V_\text{s}\Delta p}{P_\text{500}\cdot(A/\bar{d})\cdot k_\text{B}\cdot T_{500}}.
\end{equation}

\noindent With the temperature $T_{500} = 773~\text{K}$, the effective cell surface $A=2.14(10)~\text{cm}^2$, and the average wall thickness $\bar{d}= 0.10(1)~\text{cm}$, we get $\tau_\text{p}= 5.6(9)~\text{h}$, which agrees reasonably well with the measured value.  Following Swets et al. \cite{Swets61}, we can also draw conclusions about how long the gas remains in the cell if we put it, for example, in a deep freeze, namely: $\tau_\text{p}(-20^\circ\text{C})\simeq 5~\text{h}\cdot(P_{500}/P_{-20})\approx 1.6~\text{years}$ . This period can be extended even further as our sample cells are additionally sealed with a silicone-based thin film sealant (Vacseal, Space Environment Labs, Boulder, CO, USA) immediately after filling. The temperature limits for an effective seal range from liquid helium to $450^\circ\text{C}$. All cells that are not in direct use have been kept at $-20^\circ\text{C}$. The long-term behavior of such a cell based on repeated $\mathit{SNR}$ measurements is shown in Section 4.1. Let $p_\text{T}$ be the measured pressure in $V_\text{st}$ at the end of the filling process at temperature $T$ where pressure equilibrium is reached, then the room temperature (rt) pressure in the cell is given by $p_\text{rt}=p_\text{T}(T_\text{rt}/T)$ assuming the ideal gas law.\\ 

{\bf 3.2 Measures to reduce the longitudinal relaxation time $T_1$}\\

The $T_1$ relaxation time, also known as the spin-lattice relaxation time or longitudinal relaxation time, is a measure of how quickly the net magnetization vector recovers to its thermal equilibrium state in the direction of $\lvert{\bf B}\rvert$. $^3\text{He}$ is known to show very long $T_1$ relaxation times, which proves to be a drawback in its use as a magnetometer. Relaxation mechanisms include relaxation due to magnetic field inhomogeneities \cite{Schearer65}, relaxation due to paramagnetic species in the bulk gas \cite{Saam95}, and relaxation on the cell walls due to paramagnetic centers or magnetized ferromagnetic contaminants on the surface \cite{Part1, Part2, Part3}. As magnetic-field inhomogeneity induced relaxation decreases with increasing pressure, this relaxation mechanism becomes strongly suppressed in our high-pressure samples (unless this is enforced, e.g., by placing a strong permanent magnet next to the sample). At elevated pressures, the nuclear spin-relaxation rate of gaseous $^3\text{He}$ due to the magnetic-dipole interaction between the $^3\text{He}$ nuclear spins could become important: This dipolar relaxation was evaluated for temperatures from $0.1~\text{K}$ to $550~\text{K}$ \cite{Newbury93} and it can be deduced that even in the worst-case scenario for pressures of $100~\text{bar}$ and very low temperatures, the bulk dipolar relaxation in $^3\text{He}$ is still about an hour. So, again, it can be neglected for further examination. Therefore, the only option is to add paramagnetic substances and/or to substantially increase the surface-to-volume ratio $A/V$, since the wall relaxation rate is given by 
\begin{equation} \label{equ:8} 
\frac{1}{T_1^{\text{wall}}}=\rho\cdot\frac{A}{V},
\end{equation}

\noindent with $\rho$ being the surface relaxivity.\\

Our approaches to reduce $T_1$ over wide temperature ranges are:
\begin{enumerate}[i)]
\setlength{\itemsep}{0pt}
\item adding strontium peroxide $(\text{SrO}_2)$ to release paramagnetic $\text{O}_2$ from the decomposition of peroxide;
\item filling the volume with silica gel (particle size: 40-60 $\mu$m, pore size: 60~\text{\textup{\AA}}) to increase $A/V$, and
\item additional loading of the silica gel with strongly paramagnetic $\text{Gd}^{3+}$ ions.
\end{enumerate}
\bigskip

ad i) Following B. Saam et al. \cite{Saam95}, they obtained an empirical formula for the longitudinal relaxation rate  of $^3\text{He}$  in the presence of $\text{O}_2$ for $200~\text{K}\leq T\leq400~\text{K}$: 
\begin{equation} \label{equ:9} 
\Gamma_\text{O2}= 0.45[\text{O}_2](299/T)^{0.42}~\text{s}^{-1}/\text{amagat},
\end{equation}

\noindent where $[\text{O}_2]$ is the oxygen density in amagats (an amagat is a unit of density corresponding to 1~atm at $0^\circ\text{C}$).  That means, about 1~bar of $\text{O}_2$ at room temperature reduces the $T_1$ time to $\simeq 2~\text{s}$. The way to get considerable amount $>1~\text{bar}$ of $\text{O}_2$ inside the sample vessel is to use the thermal decomposition of $\text{SrO}_2$ which was added prior to sealing the cell off. The decomposition occurs with the filling procedure of $^3\text{He}$ at $\simeq 500^\circ\text{C}$ through $\text{2SrO}_2~\overrightarrow{~~_{500^\circ\text{C}}~~}~\text{2SrO} +\text{O}_2$, as discussed and investigated, e.g., in Ref.~\cite{Tribelhorn55}.  The oxygen generated remains completely in the quartz cell as its permeation rate is approximately six orders of magnitude lower than the corresponding $P_{500}$-value for helium at this temperature \cite{Norton61, Shelby97}. Therefore, any desired oxygen pressure can be set with the addition of the appropriate amount of $\text{SrO}_2$. We added 6 mg of $\text{SrO}_2$ powder inside the cell. For a stoichiometric reaction the 6~mg will generate 5 bar $\text{O}_2$ in cell volume $V_\text{s}$. Thus, we expect $T_1$~times of $\simeq 0.4~\text{s}$ at least in the temperature range where oxygen is still gaseous.\\ 

ad ii) Silica gel is an amorphous and porous form of $\text{SiO}_2$, consisting of an irregular tridimensional framework of alternating silicon and oxygen atoms with nanometer-scale voids and pores. The high specific surface area of the silica gel (300-750 $\text{m}^2/\text{g}$, \cite{Christy08}) allows it to easily absorb water, making it useful as a desiccant (drying agent). The silica gel we used for our investigations (Merck quality 60) has a pore size of 6 nm and a particle size of 40 to 63 $\mu$m. The silica gel powder was filled into a 10 cm long round bottom quartz tube to a height of $\simeq$~30~mm, i.e., about 10~mm above the sealing neck. The lower part with the silica gel was carefully heated with an initially weak flame to allow water vapor to escape without the fine powder being swept away. We increased the flame intensity until no more bubbling of the powder could be observed as a result of the escaping water vapor. The cell could then be sealed off without allowing additional vapor pressure to build up, which would make vacuum-tight sealing difficult. The remaining atmospheric oxygen [$p_\text{RT}\approx 0.2~\text{bar}\cdot(300~\text{K}/1200~\text{K})$] is essentially adsorbed in the bulk material. The adsorption equilibria and kinetics of the silica gel for $\text{O}_2, \text{N}_2$, and other gases can be found in~\cite{Park20}. \\

ad iii) The intercalation of gadolinium ions ($\text{Gd}^{3+}$) in the silica gel is achieved via a multistage process. Firstly, ca. 2~g of gadolinium(III) nitrate hexahydrate ($\text{Gd(NO}_3)_3\cdot\text{6H}_2\text{O}$) is dissolved in 10~mL of high-purity ethanol. Dried molar sieve was added to remove residual water from the solution. In parallel, the silica gel ($\simeq2$~g) was filled into a round bottom Pyrex tube with a glass flange so that it could be baked out under vacuum at $\text{300}^\circ\text{C}$ for about 5 hours. This ‘activation’ process of silica gel decreases the water content of the gel and greatly increases its adsorptive capacity \cite{Bartell32}. Finally, $\text{Gd}^\text{3+}$ dissolved in ethanol is added in an argon atmosphere of ambient pressure and temperature and left to react overnight. After removing the solvent, the silica gel was dried in a vacuum oven and the required quantity was filled into the round-bottom quartz tube with sealing neck. From then on, the steps to seal off the sample cell are the same as described in ad ii).\\

{\bf 3.3 $^3\text{He}$ NMR sample cells and NMR magnets}\\

A total of seven gas samples were investigated. The different filling pressures of $^3\text{He}$ at room temperature and the additional contents are listed in Table \ref{tab:1}. The given pressure values for the cells filled with silica gel have to be reduced by  $\simeq 26\%$ in order to compare the measured $\mathit{SNR}s$ with each other, since we have: $\mathit{SNR}\propto p\cdot V_\text{s,eff}$ [Eq. \ref{equ:3}]. The effective sample volume $V_\text{s,eff}$ was determined from the measured ratio $\mathit{SNR}_\text{30bar}^{\#4}/\mathit{SNR}_{16bar}^{\#6}=2.52$ at room temperature (see Table \ref{tab:1}). Hence, the eigenvolume of the silica gel is approximately 26\% of $V_\text{s}$.  
\begin{table}
\caption{$^3\text{He}$ sample cells used in this work with details on their respective filling pressures at RT and  volume ratios $V_\text{s,eff}/V_\text{s}$. The inner   cell volumes (empty cell), $V_\text{s}$, are identical within the specified error limits: $V_\text{s}=0.126(10)~\text{cm}^3$ .}
\begin{ruledtabular}
\begin{tabular}{lcccc}
cell&$p_\text{He}/bar$&$p_\text{O2}/bar$&silica gel&$V_\text{s,eff}/V_\text{s}$\\
\hline
\#1&$0.7$&5&-&$1$\\
	      \#2&$5$&5&-&$1$\\
          \#3&$17.2$&5&-&$1$\\
          \#4&$30$&5&-&$1$\\
          \#5&$18$&-&-&$1$\\
          \#6&$16$&-&\ding{51}&$\simeq 0.74$\\
          \#7&$16$&-&\ding{51}+$\text{Gd}^{3+}$&$\simeq 0.74$\\

\end{tabular}
\end{ruledtabular}
\label{tab:1}
\end{table}

\begin{figure}
    \centering
    \includegraphics[width=0.38\textwidth]{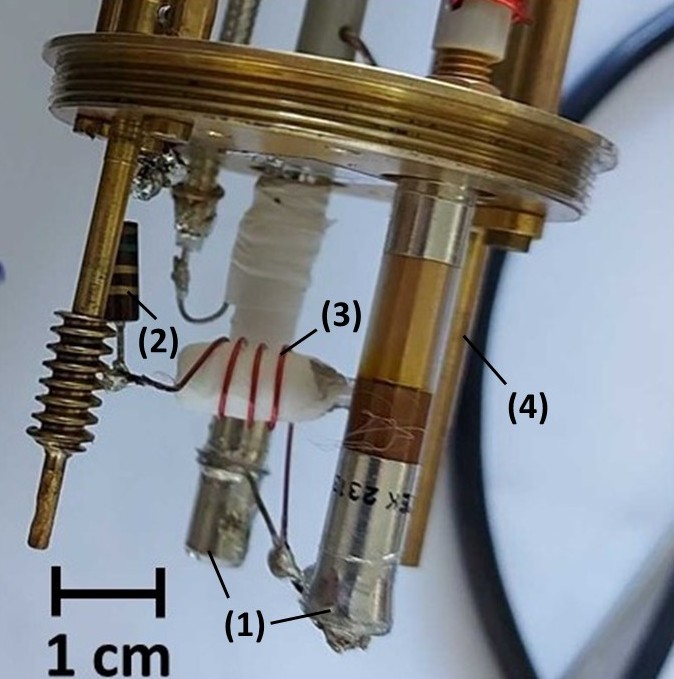}
    \caption{Probe head showing the tuning and matching capacitors (1) and the 1.5 $\Omega$ resistor (2) of the impedance-matched resonance circuit with its receiver/transmitter coil (3) wound around the sample cell under investigation. The calibrated cryogenic temperature sensor (Cernox) is placed inside the brass tube (4) next to the sample.} 
    \label{fig:4}
\end{figure}

The room temperature measurements were performed at the 7.05 Tesla wide-bore superconducting magnet (Oxford Instruments) with a Bruker Avance II spectrometer at the Max-Planck-Institute of Polymer Research (MPI-P), Mainz. The relative inhomogeneity of this magnet across the sample volume is $\Delta B/B\simeq  0.2~\text{ppm}$ which was indirectly determined from the characteristic time constant $T_2^*$  (see section 4.1) of the measured FID using Eq. \ref{equ:2}.  The quality factor $Q$ of the impedance-matched NMR circuit was determined by a network analyzer to be $Q$~=~161. Here we refer to the $Q$-value definition in Ref.~\cite{Doty88}, i.e., $Q=(\omega_0/2\pi)/\Delta f_\text{7dB}$, where $\Delta f_\text{7dB}$ is the bandwidth for which the reflected power of the matched circuit is 7~dB below the incident power. 
At the Leibniz Institute for Solid State and Materials Research (IFW Dresden), the measurements at cryogenic temperatures (to be more precise: $4.2~\text{K}\leq T < 300~\text{K}$) were carried out using their 7.05 Tesla wide-bore magnet (Oxford Instruments) together with a NMR spectrometer (Redstone, Tecmag) and a 1 kW rf-amplifier (Dressler LPPA3008). The $^3\text{He}$ samples were mounted inside a Janis STVP-NMR continuous flow cryostat cooled by $^4\text{He}$ (boiling point 4.2~K) and the temperature was regulated by a heating system (Lakeshore Model 335, temperature controller). Figure \ref{fig:4} shows the very end of the $\approx 120~\text{cm}$ long sample stick with the NMR probe head. The $Q$-value ($Q$ = 46) of the resonant circuit was spoiled by a 1.5 $\Omega$ carbon composition resistor with negative temperature coefficient ($\alpha\approx-200~\text{ppm}/^\circ\text{C}$) to reduce effects of a temperature dependent $Q$ factor on the $\mathit{SNR}$ intensity [see Eq. \ref{equ:3}]. The relative field inhomogeneity at the nominal position of the sample cell inside the superconducting magnet with $\Delta B/B\simeq 0.7~\text{ppm}$  could be deduced from the change of the measured Larmor frequencies when the sample stick was lifted up in steps of 5~mm from the nominal position of the cell.\\

{\Large\bf 4.~Experimental results}\\

{\bf 4.1 FID signals and $\mathit{SNR}$}\\

\begin{figure}
    \centering
    \includegraphics[width=0.48\textwidth]{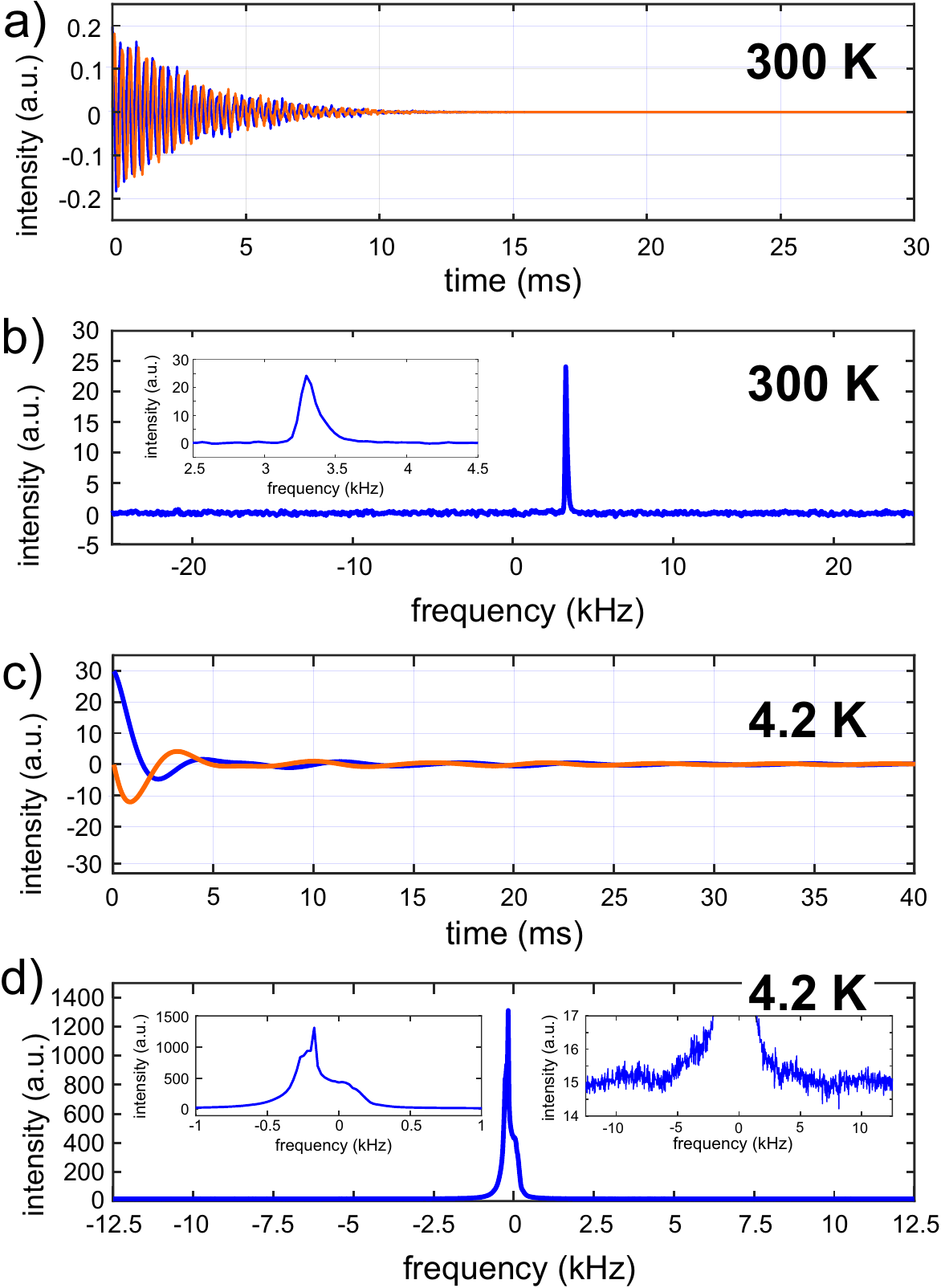}
    \caption {Free induction decay of the NMR spin precession signal and its Fourier transform measured at room temperature (MPI-P Mainz) and at 4.2~K (IFW Dresden) using a sample cell filled with 30 bar $^3\text{He}$. Paramagnetic oxygen (5 bar) was added to reduce the $T_1$-time of the thermally polarized $^3\text{He}$ gas. a) and c) show the real (blue) and imaginary (red) part of the NMR spin precession signal at 7.05~T, mixed down from 228.628 MHz to the audio-frequency range. Fourier transformation gives the spectrum of which only the real part (blue) is shown with a phase corresponding to an absorption-mode line shape. The dispersion mode (imaginary part) is not shown. The inset in b) is a zoom in the signal region to show that the line shape to a good approximation shows a Lorentzian profile, whereas the left inset in d) exhibits a spectral peak with shoulders due to first and higher order magnetic field gradient distortions. The right inset is a vertical zoom (note the intensity scale) of the entire spectrum to display the otherwise invisible noise floor. See text for more details. } 
    \label{fig:5}
\end{figure}

Figure \ref{fig:5} shows the FID signals of the 30 bar cell ($\#$4, Table \ref{tab:1}) measured at the upper and lower bounds of the investigated temperature range (MPI-P Mainz: $\simeq 300~\text{K}$ and IFW Dresden: 4.2~K). In each case, the sample magnetization was in thermal equilibrium and a resonant $\pi$/2-pulse was applied to tip it perpendicular to the magnetic field direction. The free induction decay signals (real and imaginary part) at the $^3\text{He}$ spin precession frequency of $\omega_0/2\pi = 228.628~\text{MHz}$ are mixed down to $< 4~\text{kHz}$ and are recorded with a sampling time of $\Delta t=1/f_\text{BW}= 20~\mu\text{s}$ (MPI-P) and $40~\mu\text{s}$ (IFW), respectively, corresponding to a bandwidth, $f_\text{BW}$, of 50~kHz and 25~kHz.  The chosen acquisition time was $t_\text{acq}^\text{MPI-P} = 30~\text{ms}$ at the MPI-P and $t_\text{acq}^\text{IFW} = 40~\text{ms}$ at the IFW.  For a given sample in a perfectly uniform magnetic field, the FID should be a monotonic function decreasing in time with $T_2$ [see Eq. \ref{equ:2}]. However, the FID envelope corresponding to suboptimal magnetic field homogeneity conditions may differ substantially from a pure exponential function. During our investigations, we did not put too much effort into homogenizing the field across the sample volume by optimizing the currents on the shim coils. Particularly in the temperature-dependent experiments at IFW Dresden, the homogeneity changed slightly with temperature because of thermal expansion of the sample stick moving the sample into different regions inside the magnet and the condensation of paramagnetic oxygen which generates additional local field inhomogeneities. At MPI-P Mainz an almost exponential decay of the FID is observed with $T_2^*=(3.4\pm 0.1)~\text{ms}$ . A Fourier transform of this oscillating signal shows a sharp peak at the spin precession frequency (Fig. \ref{fig:5}b), with an $\mathit{SNR}$ of 70. The width at half-maximum (FWHM) of this resonance is about 110 Hz and is somewhat broader than $\Delta\nu=1/(\pi T_2^*)=(93\pm3)~\text{Hz}$ for a Lorentzian line shape expected in the case of a purely exponential decay of the time-domain NMR signal (quadrature detection). The $\mathit{SNR}$ of all the data presented was obtained from the Fourier transform of the FID (see Appendix A). This was necessary because some of the data, such as the one shown in Fig. \ref{fig:5}c, turned out to have such a high $\mathit{SNR}$ that it was difficult to separate the noise from the signal within the chosen time window. In the frequency domain, one has a better separation of the spectral peak (Fig. \ref{fig:5}d) and the noise floor can be determined after fitting and subtracting an offset baseline. The low temperature NMR signal (Fig. \ref{fig:5}c) and its Fourier transform (Fig. \ref{fig:5}d) are also examples of signal distortions produced by first- and higher-order magnetic field gradients across the sample due to insufficient shimming and magnetic susceptibility mismatch \cite{Zanche08,Becker22}. Thus, the characteristic decay time for the FID can only be specified with a larger error, namely $T_2^*=(2.2\pm0.7)~\text{ms}$. The extracted $\mathit{SNR}$ is about 3320.  

\begin{figure}
    \centering
    \includegraphics[width=0.48\textwidth]{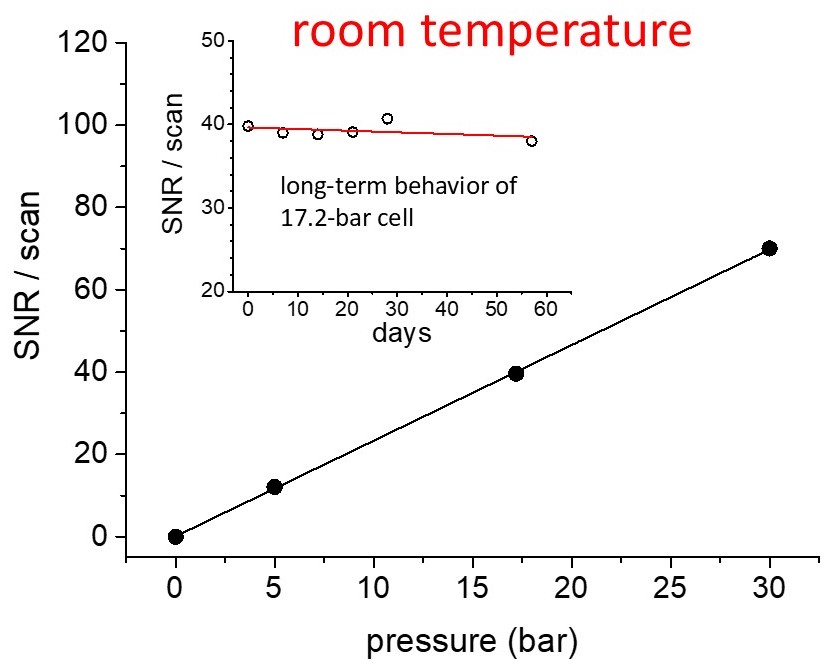}
    \caption {Pressure dependence of the $\mathit{SNR}$ measured after a $\pi/2$ pulse. The NMR samples were thermally polarized $(\simeq~300\text{K})$ in the field of the 7.05~T magnet at MPI-P, Mainz. Inset: reproducibility and long-term behavior of $\mathit{SNR}$ values, measured with the 17.2~bar cell ($\#3$, Table \ref{tab:1}).  A linear fit (solid curve) to the $\mathit{SNR}$-data points results in a slope $m = -0.020(21)~\text{day}^{-1}$, compatible with zero.  } 
    \label{fig:6}
\end{figure}

Figure \ref{fig:6} shows the dependence of the $\mathit{SNR}$ from the $^3\text{He}$ gas pressure measured at room temperature (MPI-P Mainz) with cells $\#2$, $\#3$, and $\#4$ (see Table \ref{tab:1}). The samples were again in thermal polarization equilibrium and a single resonant $\pi/2$-pulse was applied to deduce the respective $\mathit{SNR}$ per scan. The curve shows a strictly linear behavior with a slope of $m=2.32(2)~\text{bar}^\text{-1}$, so that we expect an $\mathit{SNR}~>~200$ in the accessible pressure range up to 100 bar. The reproducibility of a measured $\mathit{SNR}$ is shown using the example of the 17.2-bar cell ($\#3$), a series of measurements that extended over two months (inset of Fig. \ref{fig:6}). This also constitutes a long-term measurement that shows no obvious signal drop due to $^3\text{He}$ leakage within the fluctuations of the $\mathit{SNR}$ data points.

\begin{figure}
    \centering
    \includegraphics[width=0.48\textwidth]{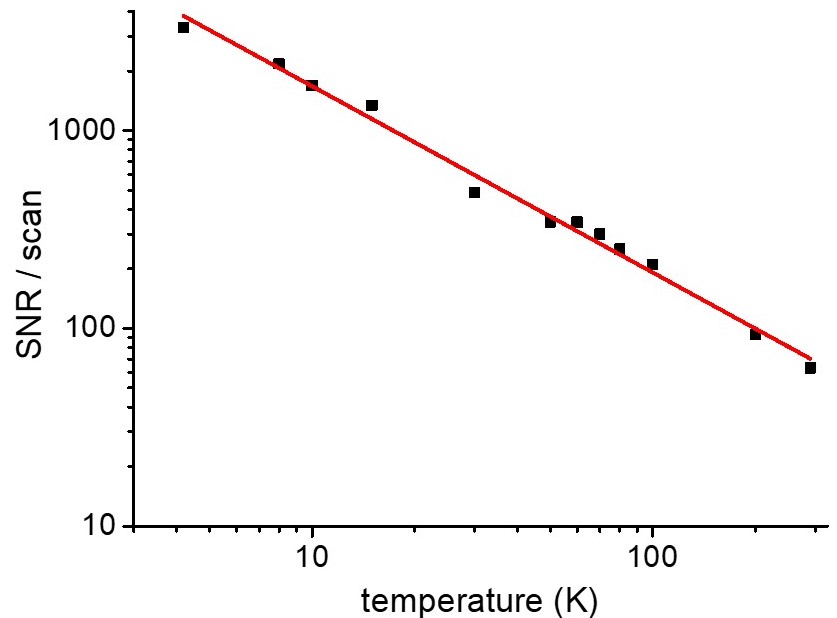}
    \caption {Log-log plot of the temperature dependence of the $\mathit{SNR}$ measured with the 30 bar cell ($\#4$, Table \ref{tab:1}) in the range $4.2~\text{K}\leq T < 300\text{K}$ at IFW Dresden. From a linear fit (red curve) to the data one deduces a  $\propto T^{-0.94(3)}\approx 1/T$ dependence of the $\mathit{SNR}$ on temperature. } 
    \label{fig:7}
\end{figure}

The temperature dependence of the measured $\mathit{SNR}$ of the 30 bar sample cell ($\#4$, Table \ref{tab:1}) is shown in Fig. \ref{fig:7}. From the fit to the data points (log-log plot) one derives a $\propto T^{-0.94(3)}$ scaling of the $\mathit{SNR}$. This is in fairly good agreement with the expected $1/T$ dependence, for which we have to assume that the quality factor $Q$ of the NMR resonance circuit does not depend on temperature [see Eq. \ref{equ:3}], an assumption that is only fulfilled to a first approximation. At room temperature, the measured $\mathit{SNR}$ values at MPI-P and IFW can be compared giving $R=\mathit{SNR}^\text{MPI-P}/\mathit{SNR}^\text{IFW}=70/63=1.11$. Considering the $\propto \sqrt{Q/f_\text{BW}}$ dependence of the $\mathit{SNR}$ [Eq. \ref{equ:3}], the expected ratio should be $R_\text{exp}=1.32$. We explain this deviation of approx. $16\%$ by the slightly different parameters $K$ and $\eta$ for the coil geometry and the filling factor, which also influence $\mathit{SNR}$ [see Eq. \ref{equ:3}].\\

{\bf 4.2 $T_1$ and $T_2^*$ relaxation times}\\

\begin{figure}
    \centering
    \includegraphics[width=0.48\textwidth]{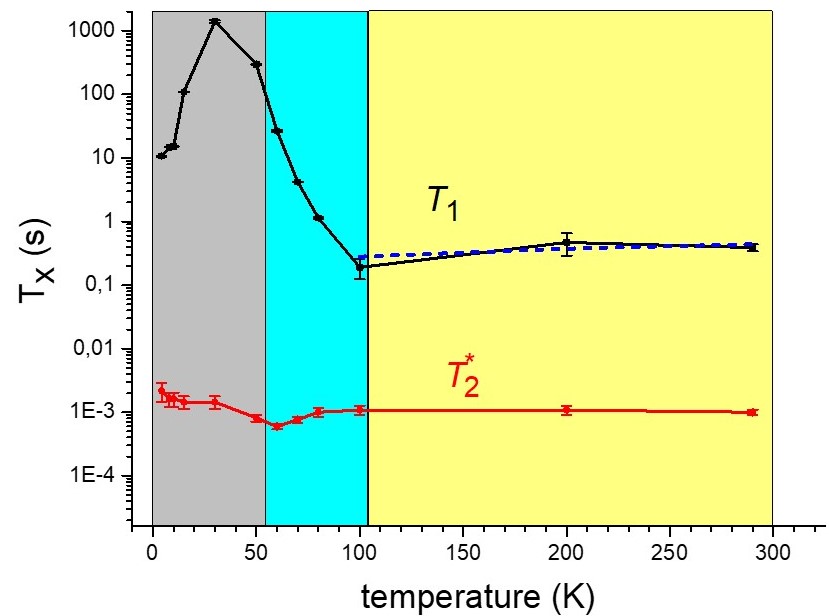}
    \caption {Temperature dependence of the longitudinal ($T_1$) and transverse ($T_2^*$) relaxation time of the 30 bar $^3\text{He}$ cell ($\#4$, Table \ref{tab:1}) in the temperature range $4.2~\text{K}\leq T < 300~\text{K}$. The three colored areas indicate the state of aggregation of the added oxygen: yellow (gaseous), light blue (liquid), and grey (solid). The blue dashed line is the expected $T_1$-relaxation of $^3\text{He}$ in presence of gaseous, paramagnetic oxygen (filling pressure: 5~bar). See text for more details. } 
    \label{fig:8}
\end{figure}

The longitudinal relaxation time $T_1$ of our $^3\text{He}$ samples is measured by the saturation recovery method \cite{Levitt08}. Saturation recovery comprises a train of NMR excitation pulses that randomize the spins of the $^3\text{He}$ atoms. Then a variable time gap $\tau$ is introduced during which magnetization recovers towards equilibrium. Finally, a $\pi/2$ pulse is applied to measure the magnitude of the NMR signal.  From the time evolution of the magnetization $M(\tau)$ given by  $M(\tau) = M_0 [1 - \text{exp}(-\tau/T_1 )]$, the $T_1$ time can be extracted. In case of the 30 bar $^3\text{He}$ cell, the temperature dependencies of the $T_1$ time and the transverse relaxation time $T_2^*$ (extracted from the measured FID) are shown in Fig. \ref{fig:8}.

A striking feature is the pronounced dependence of the $T_1$ time on the aggregate state of oxygen, as highlighted by the three distinct color-coded regions. At 5 bar oxygen filling pressure, phase transitions occur at 105~K (gaseous/liquid) and 54~K (liquid/solid). Whereas oxygen is paramagnetic in the gaseous and liquid phase with magnetic susceptibility $\chi_\text{v}\propto 1/T>0$, one can identify three crystallographic phases ($\alpha$, $\beta$, and $\gamma$) in the solid state, where the $\alpha$ and $\beta$ phase represent an antiferromagnetic order going along with a drop in the magnetic susceptibility, but still $\chi_\text{v}>0$ \cite{Gregory78,Jezowski06}. The measured $T_1$ values and their temperature dependencies in the gaseous phase can be well described (blue dashed line in Fig. \ref{fig:8}) by the empirical formula [Eq. \ref{equ:9}] from B. Saam et al. \cite{Saam95}, saying that a desired $T_1$ time can be reached by adjusting the oxygen filling pressure. Below $T=105~\text{K}$ oxygen freezes out on the cell walls and forms a thin layer of about $3.9~\mu\text{m}$ average thickness. The $T_1$ relaxation is no longer controlled by the strong gas-phase intermolecular dipolar relaxation, but is governed by the weaker wall relaxation rate according to Eq. \ref{equ:8} with the surface relaxivity, $\rho$, attributable to the paramagnetic oxygen layer. The $T_1$ time increases and reaches its maximum of $T_1^\text{max}\approx 20~\text{min}$ at around $T=30~\text{K}$, i.e., where oxygen is already in the solid state (antiferromagnetic).  Below 30~K, $T_1$ drops and reaches $T_1\approx 10~\text{s}$ at $T=4.2~\text{K}$. The decrease in $T_1$ towards lower temperatures is a result of the increase in adsorption-dominated relaxivity, since the sticking time $\tau_\text{s}$  per wall collision follows Frenkel’s law \cite{Frenkel24} with $\tau_\text{s}=\tau_\text{s,0}\cdot\text{exp}(E_\text{ad}/k_\text{B}T)$, where $E_\text{ad}$  is the adsorption energy and $\tau_\text{s,0}\approx 10^{-13}~\text{s}$. Here, the reader is advised to refer to Ref.~\cite{Part1}, where the temperature dependence of the surface relaxivity is discussed in more detail. In contrast to the $T_1$ time, the transversal relaxation time $T_2^*$ remains essentially constant (within a factor of 2) over the entire temperature range. We attribute the small changes at temperatures below 80~K to local field gradients induced by the magnetic susceptibility (temperature-dependent) of the paramagnetic/antiferromagnetic oxygen layer which change the field homogeneity across the sample volume if the sample holder does not have spherical symmetry \cite{Maul16}. The local field gradients may also be altered by the non-homogeneity of the oxygen layer, e.g., if the liquid tends to collect on the bottom of the cell. In its gaseous state, the paramagnetism of oxygen is not seen by the colocated $^3\text{He}$ spins in the case of  spherical sample geometry. Otherwise, demagnetization fields may change the local $B$-field and its homogeneity \cite{Vlassenbroek96}.  A quantitative analysis of these effects is beyond the scope of our investigations, but will be important when it comes to the absolute measurement of magnetic fields.

\begin{figure}
    \centering
    \includegraphics[width=0.48\textwidth]{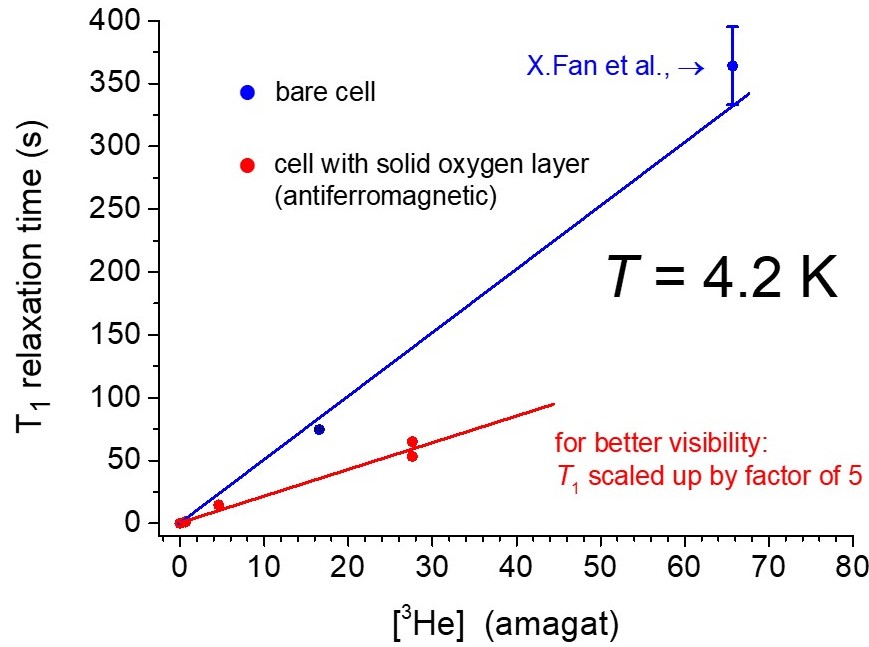}
    \caption {$T_1$ relaxation plotted versus [$^3\text{He}$] in units of amagat. Error bars in part smaller than symbol size. The lines are linear fits through the origin for bare cells (blue data points) and cells with a solid $\text{O}_2$ layer (red data points). The linear dependence of $T_1$ (wall relaxation) on [$^3\text{He}$] is expected at low temperatures such as $T = 4.2~\text{K}.$ } 
    \label{fig:9}
\end{figure}

Figure \ref{fig:9} shows the pressure dependence of the $T_1$ relaxation measured at $T=4.2~\text{K}$. Pressure values (at room temperature) are expressed in units of amagat. We observe a linear dependence for both the bare cells and the sample cells with the added oxygen (solid oxygen layer). The data of the bare cell ($\#5$) was complemented by the $T_1$ value measured on a $0.5~\text{cm}^3$ NMR sample bulb (Pyrex) with similar surface to volume ratio \cite{Fan19}.  However, the respective surface relaxivities $\rho\propto 1/T_1$ differ by a factor of about 12 as can be deduced from the ratio of the slopes from the straight-line fits to the data. A similar result in $T_1$ behavior ($^3\text{He}$ on solid oxygen and $^3\text{He}$ on bare glass) was also observed in Ref.~\cite{Lefevre88} in the temperature range $14~\text{K} < T < 50~\text{K}$ at even lower particle densities $(2.5\times 10^{-3}~\text{amagat})$. The pressure dependence of the $T_1$ time gives some control over the characteristic time to reach thermal polarization equilibrium and thus to set the optimal sensor read-out rates.  In some cases, this may be more advantageous than maximizing the already high $\mathit{SNR}$ and, with it, the associated measurement sensitivity of the $^3\text{He}$ magnetometer. The monolayer coverage, $X$, is the criterion to observe a linear relationship between the $T_1$ relaxation of the wall and the number density of particles \cite{Lusher88, Lefevre85}. A complete monolayer will be formed by 
\begin{equation} \label{equ:10} 
X=n_\text{v}\lambda_\text{th}/s_0\cdot\text{exp}(\Delta W/k_\text{B}T)\approx 1,
\end{equation}

\noindent where $s_0$ is the areal density of helium at monolayer completion $(s_0\simeq 0.1\textup{\AA}^{-2})$, $n_\text{v}$ is the particle number density in the sample volume with $n_\text{v}=6.28\times 10^{19}~[^3\text{He}]$, $\lambda_\text{th}$ is the thermal de Broglie wavelength with $\lambda_\text{th}\simeq 10 \textup{\textup{\AA}} /\sqrt{T}$ for $^3\text{He}$, and $\Delta W$ is the adatom-substrate binding energy.  In Ref.~\cite{Lefevre88}, $\Delta W$ was determined to $\Delta W/k_\text{B}(^3\text{He}/\text{O}_2)=(130\pm 15)~\text{K}$ and $\Delta W/k_\text{B}(^3\text{He}/\text{glass})=(130\pm 20)~\text{K}$, so both binding energies are equal within the scope of their errors. For the data shown in Fig. \ref{fig:9}, monolayer completion is fulfilled. In contrast, for $X\ll 1$ the $T_1$ wall relaxation time would become pressure independent.\\

\begin{figure}
    \centering
    \includegraphics[width=0.48\textwidth]{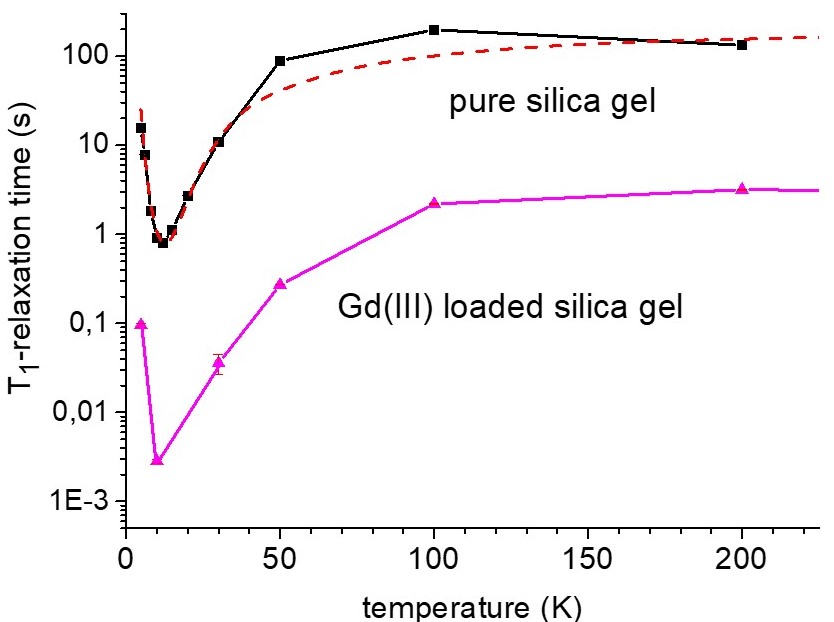}
    \caption {Temperature dependence of the characteristic $T_1$ relaxation time of thermally polarized $^3\text{He}$ for both the pure silica gel cell ($\#6$) and the Gd(III) loaded silica gel cell ($\#7$). Error bars in part smaller than symbol size. Both curves have approximately the same course with a pronounced minimum at $T=12~\text{K}$. In case of Gd(III), however, the absolute $T_1$ values are about 2 orders of magnitude lower. The red dashed line is a fit to the pure silica-gel data based on the BPP model.} 
    \label{fig:10}
\end{figure}

The measured $T_1$ relaxation times of the $^3\text{He}$ cells filled with silica gel (pure silica gel ($\# 6$) and Gd(III) loaded silica gel ($\# 7$)) show a completely different temperature dependence compared to samples containing oxygen (see Fig. \ref{fig:10}): In both cases, there is a pronounced minimum around $T=12~\text{K}$. Overall, the functional dependence on temperature is the same, but in absolute values the $T_1$ times of both samples differ by about two orders of magnitude. Depending on the Gd(III) concentration, $[\text{Gd}^{3+}]$, the desired $T_1$ times can be set, approaching the pure silica gel $T_1$-values for $[\text{Gd}^{3+}]\rightarrow 0$.
The temperature dependence of $T_1$ can be described by the Bloembergen–Purcell–Pound (BPP) model \cite{BPP48}. The simplest approach for diffusion in ordered systems assumes an exponential correlation function proportional to $\text{exp}(-t/\tau_\text{c})$ where $\tau_\text{c}$ is the mean time for one of a pair of interacting spins to jump. The corresponding spectral density function is a Lorentzian function of frequency $1/\tau_\text{c}$ . It is also assumed that $\tau_\text{c}$ depends on temperature $T$ according to the Arrhenius form $\tau_\text{c} = \tau_0\cdot\text{exp}(E_\text{a}/k_\text{B}T)$ where $E_\text{a}$ is the activation energy for the diffusive jump of a spin. In the BPP model with a single activation energy, $1/T_1$ is given by
\begin{equation} \label{equ:11} 
\frac{1}{T_1}=H\cdot\left (\frac{\tau_\text{c}}{1+\omega_0^2\tau_\text{c}^2}+\frac{4\tau_\text{c}}{1+4\omega_0^2\tau_\text{c}^2}\right ),
\end{equation}

\noindent and the $T_1$  minimum occurs for $\omega_0\tau_\text{c}=0.616$. In disordered systems the structural disorder will produce a distribution of activation energies. The common extension of the BPP model for such cases is to integrate over a Gaussian distribution of activation energies, $g(E_\text{a})= 1/(\sqrt{2\pi}\sigma)\cdot\text{exp}\left (-(E_\text{a}-\langle E_\text{a}\rangle)^2/2\sigma^2 \right )$, to get the weighted average of the ordered BPP model
\cite{McDowell01}
\begin{equation} \label{equ:12} 
\langle T_1(T)\rangle = \Big(\int_0^\infty g(E_\text{a})\cdot (\frac{1}{T_1})\,dE_\text{a} )\Big)^{-1}.
\end{equation}
The fit of $\langle T_1(T)\rangle$ to the pure silica gel data is shown in Fig. \ref{fig:10} by the red dashed line with the fit results for the free model parameters given by $\tau_0 = 4\cdot10^{-13}~\text{s}$, $\langle E_\text{a}/k_\text{B}\rangle=83.8~\text{K}$, $\sigma/k_\text{B}=20~\text{K}$ and $H=2.12\times 10^9~\text{s}^{-2}$. 

We refrained from presenting the measured temperature dependence of the $T_2^*$ transversal relaxation times for the pure silica gel sample, because it is essentially the same as for the 30 bar $^3\text{He}$ cell (see Fig. \ref{fig:8}). On the other hand, with the Gd(III) loaded silica-gel cell, the $T_2^*$  times drop significantly below $T=200\text{K}$ and reach values $< 0.1~\text{ms}$ towards 4.2~K. Especially in the lowest temperature range, the obvious high concentration of Gd(III) ions has a detrimental effect on the achievable sensitivity for the frequency estimation $\sigma_f\propto 1/(T_2^*)^{3/2}$ and thus for the accuracy of the $B$ field measurement. \\

{\bf 4.3 Sensitivity in magnetic field measurement}\\

\begin{figure}
    \centering
    \includegraphics[width=0.48\textwidth]{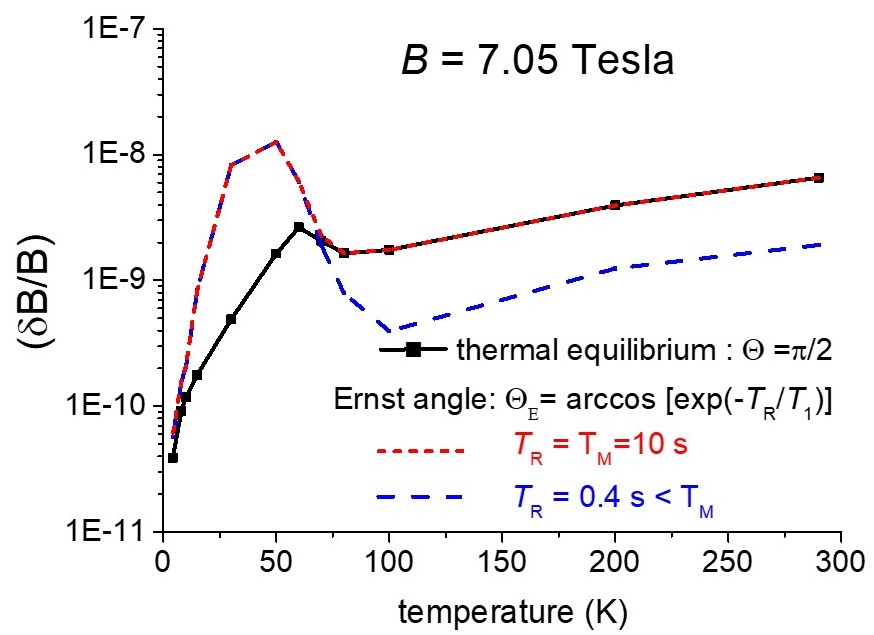}
    \caption {Relative sensitivity $(\delta B/B)$ vs temperature. Highest sensitivity per NMR scan (black squares) for $^3\text{He}$ sample (30 bar) in thermal polarization equilibrium ($\pi/2$ flip). Sensitivity per NMR scan in case of field monitoring with repetition time  $T_\text{R} =10~\text{s}$ (red short-dashed curve) under the optimal flip angle $\Theta_\text{E}$ (Ernst angle). The input parameters for the CRLB magnetic field estimation [Eqs. \ref{equ:5}, \ref{equ:6}] can be found in the text and be taken from the relevant figures (Fig. \ref{fig:7}, Fig. \ref{fig:8}). Improvement of measurement sensitivity  per unit time $(T_\text{m}=10~\text{s})$ by signal accumulation with a repetition time of $T_\text{R}=0.4~\text{s}$, in particular in regions of the sample's $T_1$ times where $T_1 < T_\text{m},~ T_\text{R}$ (blue dashed curve ).    } 
    \label{fig:11}
\end{figure}

The sensitivity of the cryogenic $^3\text{He}$ magnetometer can be determined using the available data. Measurement accuracy means, that we are talking about the relative accuracy in the determination of the $B$-field. For an absolute determination of the magnetic field, further requirements must be met, which were not the focus of our investigations. In particular, the geometry of the sample container is decisive. The suppression of susceptibility artifacts can best be achieved with spherical cells, as in the case of a gaseous NMR sample \cite{Maul16}.  The influence of the sample environment on the local field across the sample must also be taken into account and is discussed in detail, e.g. in Refs.\cite{Fei97, Zanche08}. These measures would be the logical steps in a further stage of development of the cryogenic $^3\text{He}$ magnetometer toward a sensor for absolute magnetic field measurements.  In order to determine the accuracy limits of a relative measurement of magnetic fields, the CRLB formula [Eq. \ref{equ:5}] will be utilized. We use the data from the 30 bar $^3\text{He}$ cell, as these measurements were carried out at the highest spin density and therefore the highest $\mathit{SNR}$ values. In Fig. \ref{fig:11}, $\delta B/B$ (black data points) is shown as a function of temperature for single-pulse NMR measurements with the $^3\text{He}$ spin sample in thermal polarization equilibrium. The $\pi/2$ flip angle gives the maximal NMR signal in case of $T_\text{R} \gg T_1$, where $T_\text{R}$ is the repetition time until the next excitation pulse. For long $T_1$ samples, e.g., $T_1\simeq 20~\text{min}$ in $T=30~\text{K}$ (see Fig. \ref{fig:8}), the sensor read-out rates become very moderate. \\
\noindent If the magnetometer is to be operated in field-monitor mode for pulse sequences with a recovery time of $T_\text{R} < T_1$, this does not allow the spin system to completely relax back to equilibrium. The flip angle that maximizes the magnetization in the transverse plane, $M_\perp$, and thus gives the maximal signal intensity is called the Ernst angle, $\Theta_\text{E}$, which can be calculated from the equation  $\Theta_\text{E}=\arccos~[\text{exp}(-T_\text{R}/T_1)]$ \cite{Ernst66} and $M_\perp$ is given by
\begin{equation} \label{equ:13} 
M_\perp=M_\text{ss}\cdot\sin\Theta_\text{E}=M_0 \frac{1-\text{exp}(-T_\text{R}/T_1)}{\sqrt{1-\text{exp}(-2T_\text{R}/T_1)}}~,
\end{equation}
where $M_\text{ss}$ is the available longitudinal steady state magnetization. For $T_\text{R} \gg T_1$, the transverse magnetization reaches its maximum value, $M_\perp=M_0$, as the excitation angle $\Theta_\text{E}$ asymptotically approaches $\pi/2$. 
When $T_\text{R} < T_1$, $M_\perp$ can be approximated as $M_\perp\simeq M_0\sqrt{T_\text{R}/2T_1}$ .
The magnetometer sensitivities $\delta B/B$ for a chosen sensor read out rate of $1/T_\text{R}=0.1~\text{Hz}$  are shown in Fig. \ref{fig:11}. These are represented by the red short-dashed curve, which is based on the $T_1$ data shown in Fig. \ref{fig:8}. In the temperature range $100~\text{K} < T < 300~\text{K}$, the accuracy of field measurements remains largely umaffected, as 
$T_\text{R}>T_1$. A similar trend is observed when approaching $4.2~\text{K}$, where $T_1$ approaches $T_\text{R}$. However, in the temperature range around 30 K, a significant decrease in measurement sensitivity is evident due to $T_1\gg T_\text{R}$. \\
\noindent With the accumulation of $n$ NMR scans and assuming the random noise to be white, the $\mathit{SNR}$ scales with $\sqrt{n}$ and so does the magnetometer sensitivity. Monitoring magnetic field changes, the maximum $\mathit{SNR}$ per unit time $(T_\text{m})$ is the decisive quantity \cite{RErnst66}. In certain cases, accumulating NMR scans can significantly enhance the sensitivity of magnetic field measurements. 
For $n=T_\text{m}/T_\text{R}$, $\mathit{SNR(n)}$ is proportional to
\begin{equation} \label{equ:14} 
SNR(n)\propto \sqrt{\frac{T_\text{m}}{T_1}}\frac{1-\text{exp}(-T_\text{R}/T_1)}{\sqrt{(T_\text{R}/T_1)(1-\text{exp}(-2T_\text{R}/T_1))}}.
\end{equation}

\noindent Applied to our example above, where we had used $T_\text{m}=T_\text{R}=10~\text{s}~(n=1)$, there is a significant increase in measurement sensitivity for $T_\text{m}=10~\text{s}$ and $T_\text{R}= 0.4~\text{s}~(n=25)$ in $T_1$ regions where $ T_1 < T_\text{m},~T_\text{R}$ (blue dashed curve in Fig. \ref {fig:11}). On the other hand, the accumulation of NMR scans has little effect on the measurement sensitivity $\delta B/B$ for $T_1 > T_\text{m}$. 

In the following, all characteristic parameters that are relevant for the investigations carried out in the 7.05 Tesla field or could be determined therefrom are labeled with the index ‘$_0$’. They will be used as input parameters to obtain reliable information on $\delta B/B$ and $\mathit{SNR}$ for other $B$-fields, with a focus on lower magnetic fields to examine the range limit down to which the $^3\text{He}$ magnetometer can be satisfactorily operated. However, some simplified but realistic assumptions are necessary in order to explore the accessible field range of this sensor: a) the $Q$-value of the NMR resonance circuit is constant and we set $Q(B,T)=100$ ($Q_0=46$); b) $T_2^*$ is independent of temperature (see Fig. \ref{fig:8}) and its magnetic field dependence is $T_2^*=(T_2^*)_0\cdot(B_0/B)$.
The latter follows from Eq. \ref{equ:2} assuming a constant relative field inhomogeneity of 0.7 ppm with $(T_2^*)_0=1~\text{ms}$; c) $\mathit{SNR}$ shows a $1/T$-dependence and scales with $B^2$. The quadratic dependence on $B$ results from Eq. \ref{equ:3}, being aware that in the derivation of this equation \cite{Hoult76} the quality factor $Q$ exhibits a linear dependence on $\omega_0$; d) as the $\mathit{SNR}$ increases linearly with the $^3\text{He}$ pressure (Fig. \ref{fig:6}), we take $p=80~\text{bar}$ for the computation of the sensor's sensitivity limits instead of $p_0=30~\text{bar}$ and $\mathit{SNR}_0$ refers to the measured value at $T=300~\text{K}$, i.e., $\mathit{SNR}_0=63$; e) the bandwidth $(f_\text{BW})_0$ of 25 kHz is adopted, although this parameter can also be fine-tuned. The magnetic field and temperature dependence of the $\mathit{SNR}$ can then we written
\begin{equation} \label{equ:16} 
\mathit{SNR}=\mathit{SNR}_0\Big(\frac{B}{B_0}\Big)^2\Big(\frac{300~\text{K}}{T}\Big)\sqrt{\frac{Q}{Q_0}}\frac{p}{p_0}.
\end{equation}

\begin{figure}
    \centering
    \includegraphics[width=0.48\textwidth]{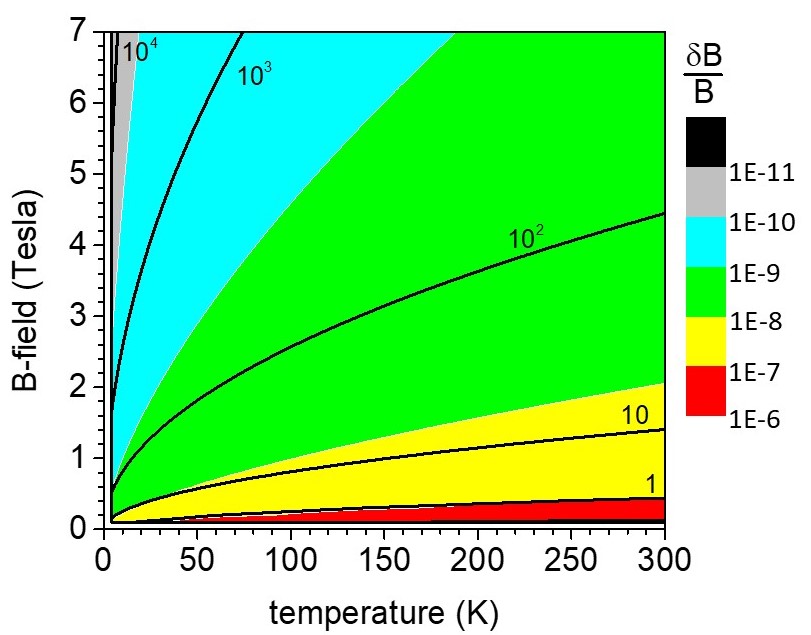}
    \caption {Contourplot of the relative accuracy $(\delta B/B)$ of magnetic field measurements in the $B-T$ plane . The black solid lines give the $\mathit{SNR}$ values per NMR scan ($\pi/2$ flip) graded by magnitude. } 
    \label{fig:12}
\end{figure}

Figure \ref{fig:12} shows the expected sensitivities $\delta B/B$ of the $^3\text{He}$ magnetometer as well as the corresponding $\mathit{SNR}$-values, both graded in magnitudes across the respective magnetic field- and temperature ranges $0.1~\text{T} < B\leq 7~\text{T}$ and $4~K < T \leq 300~\text{K}$. For $\mathit{SNR} > 1$, the sensitivity limits that cover the range from $10^{-11} < (\delta B/B) < 10^{-7}$, are accessible in a single pulse ($\pi/2$)-NMR measurement with the sample in thermal polarization equilibrium. The accumulation of NMR scans also allows access to the edges of sensor operation, i.e. $T\rightarrow 300~\text{K}$ and low magnetic fields $(\approx 0.1~\text{T})$. \\

{\bf\Large 5. Conclusion}\\

In this paper, we have presented a sensitive $^3\text{He}$ magnetometer to monitor magnetic fields of $B > 0.1~\text{T}$  in an environment from ambient temperatures down to $4~\text{K}$. Our approach is based on the NMR measurement of the free induction decay of thermally polarized $^3\text{He}$ after a resonant radio frequency pulse excitation. In order to reach a high spin density, we use the selective permeation of helium through quartz to fill small sample cells with a helium pressure up to $100~\text{bar}$. The otherwise long $T_1$-relaxation times of minutes to hours to reach thermal polarization equilibrium could be reduced to $\mathcal{O}$ (s) over a wide temperature range that allows field monitoring with high sensor read-out rates.
The demonstrated approaches to reduce $T_1$ exhibit varying effectiveness across the investigated temperature range. Researchers are advised to carefully select and tailor the setup of the $^3\text{He}$-magnetometer to the required operating temperature. The compact, reliable, and easy to handle $^3\text{He}$ NMR probes are ideal for shimming and measuring the stability of superconducting solenoids with cryogenic bores. The flexibility in positioning the probe enables the acquisition of magnetic field maps. \\

{\bf Acknowledgments}\\
The authors are grateful to glass blower A. Schwaitzer and D. Schellen from the Institute of Technology and Engineering (ITE) at the RC J\"ulich for preparing our quartz cells. The authors also thank H.-J. Krause from Institute of Biological Information Processing (IBI) at RC J\"ulich for coordinating the glass technical orders. V. Denysenkov and T. Prisner from the Institute for Physical and Theoretical Chemistry, Goethe University Frankfurt have to be acknowledged for their assistance in the initial phase of this project. The authors are grateful to Beate Müller from the Max-Planck Institute of Polymer Research, Mainz, Germany, for her expert guidance on silica gels. This work was supported by the Deutsche Forschungsgemeinschaft (DFG) under contract number 542505174.\\

{\bf\large Appendix A: \textit{SNR} Determination}\\

\renewcommand{\theequation}{A.\arabic{equation}}
\setcounter{equation}{0}
The basis of Eq. \ref{equ:5} is an exponentially damped sinusoidal signal of frequency $\omega_0$  with a characteristic decay time $T_2^*$ . The signal strength is given by the amplitude $|S(t=0)|=S_0$ and the noise level ($noise$) 
\begin{equation} \label{equA:1} 
S(t)=S_0\cdot\text{exp}(i\omega_0t-t/T_2^*+i\varphi)+noise.
\end{equation}

\noindent This is a complex signal as typically detected in NMR experiments to allow for a sign-specific Fourier transformation. The NMR radio-frequency signal is reduced to the audio-frequency range by mixing out the high frequency component: $\Delta\omega_0=\omega_0-\omega_\text{ref}$ . The low-frequency audio spectrum is then digitized by $N$ (complex) points sampled at equal time intervals $\Delta t$ and properly phase corrected ($\varphi = 0$), so that $\Re(S(0))=S_0$ and $\Im(S(0))=0$. The recorded FID is given by
\begin{align} \label{equA:2}
S_k=S(k\Delta t)=&S_0\cdot\text{exp}\Big((i\Delta\omega_0-1/T_2^*)k\Delta t\Big)+ noise_k\nonumber\\
&\text{with}~ k = 0, 1, 2,..., N-1,
\end{align}

\noindent and with $N=t_\text{acq}/\Delta t$. The discrete Fourier transform (DFT) transforms the  sequence of $N$ complex numbers $\{S_k\}:=S_0,S_1,...,S_{N-1}$ into another sequence of complex numbers, which is defined by  
\begin{align} \label{equA:3} 
F_n=&\frac{1}{\sqrt{N}}\sum_{k=0}^{N-1} S_k\cdot\text{exp}(-i\omega_nk\Delta t)\nonumber\\
&\text{for} \quad -(N-1)/2\leq n \leq (N-1)/2,
\end{align}

\noindent with $\omega_n=(n\cdot\Delta\omega)$. Here we used the symmetrical definition of the DFT as implemented by the used software (GNU Octave, ver. 8.4.0). The discrete Fourier transform of the digitized FID data produces a spectrum of $N$ complex points at frequency intervals $\Delta\omega=1/(N\Delta t)$. Since we must refer to the real value in the time domain $\mathit{SNR}$ as used in Eq. \ref{equ:5}, the $\mathit{SNR}$ in the frequency domain is extracted from the real part of the DFT given by 
\begin{equation} \label{equA:4} 
\mathit{SNR}=\frac{1}{\sqrt{N}}\frac{\sum_{signal(n)}\Re (F_n)}{\sqrt{\frac{1}{M}\sum_{noise(n)}\big(\Re(F_n)\big)^2}}.
\end{equation}

\noindent This means, that we have to sum over the signal carrying part of the spectrum ($signal(n)$:$\approx 5\times$~linewidth of the spectral peak) and to divide by the standard deviation of the noisy region $noise(n)$ of the spectrum with a mean of zero (after subtraction of an offset baseline).  $M$ denotes the number of data points in the noise data set. 


\end{document}